\documentclass[article,twocolumn,showpacs,preprintnumbers,amsmath,amssymb, superscriptaddress]{revtex4-1}
\usepackage{graphicx}
\usepackage{dcolumn}
\usepackage{bm}
\usepackage{fancyhdr}
\usepackage{float}
\usepackage{ifthen}
\usepackage{eurosym}
\usepackage{wrapfig}
\usepackage{xfrac}

\begin{document}
\title{Distinct domain switching in Nd$_{0.05}$Ce$_{0.95}$CoIn$_5$ at low and high fields}

\ \author{D.~G.~Mazzone}
\altaffiliation{danimazzone@gmail.com}
\ \affiliation{Laboratory for Neutron Scattering and Imaging, Paul Scherrer Institut, 5232 Villigen PSI, Switzerland}

\ \author{R. Yadav}
\ \affiliation{Laboratory for Scientific Developments and Novel Materials, Paul Scherrer Institut, 5232 Villigen PSI, Switzerland}

\ \author{M. Bartkowiak}
\ \affiliation{Laboratory for Scientific Developments and Novel Materials, Paul Scherrer Institut, 5232 Villigen PSI, Switzerland}

\ \author{J.~L.~Gavilano}
\ \affiliation{Laboratory for Neutron Scattering and Imaging, Paul Scherrer Institut, 5232 Villigen PSI, Switzerland}

\ \author{S.~Raymond}
\ \affiliation{Univ. Grenoble Alpes and CEA, INAC, MEM, F-38000 Grenoble, France}

 \author{E. Ressouche}
\ \affiliation{Univ. Grenoble Alpes and CEA, INAC, MEM, F-38000 Grenoble, France}

\ \author{G.~Lapertot}
\ \affiliation{Univ. Grenoble Alpes and CEA, INAC, PHELIQS, F-38000 Grenoble, France}

\ \author{M.~Kenzelmann}
\ \affiliation{Laboratory for Scientific Developments and Novel Materials, Paul Scherrer Institut, 5232 Villigen PSI, Switzerland}
\ \affiliation{Laboratory for Neutron Scattering and Imaging, Paul Scherrer Institut, 5232 Villigen PSI, Switzerland}

\date{\today}
\begin{abstract}
Nd$_{0.05}$Ce$_{0.95}$CoIn$_5$ features a magnetic field-driven quantum phase transition that separates two antiferromagnetic phases with an identical magnetic structure inside the superconducting condensate. Using neutron diffraction we demonstrate that the population of the two magnetic domains in the two phases is affected differently by the rotation of the magnetic field in the tetragonal basal plane. In the low-field SDW-phase the domain population is only weakly affected while in the high-field Q-phase they undergo a sharp switch for fields around the $a$-axis. Our results provide evidence that the anisotropic spin susceptibility in both phases arises ultimately from spin-orbit interactions but are qualitatively different in the two phases. This provides evidence that the electronic structure is changed at the quantum phase transition, which yields a modified coupling between magnetism and superconductivity in the Q-phase. 
\end{abstract}
\maketitle
\section*{Introduction}

Strongly correlated electron systems can feature electronic ground states, in which different electronic charge, spin, orbital and lattice degrees of freedom such as phonons, defects or strain are coupled. Such couplings can trigger novel quantum phenomena, such as unconventional superconductivity, topological and multiferroic phases or heavy-fermion ground states. The understanding of cooperative phenomena is particularly challenging in unconventional superconductors where Cooper pairs are thought to arise from magnetic fluctuations \cite{Monthoux2007}. Antisymmetric spin-orbit interactions can lead to novel phases with uncommon symmetry such as triplet superconductivity.  In CePt$_3$Si, for instance, it is believed that such interactions generate an anomalous spin susceptibility that triggers superconductivity with mixed spin-singlet and triplet Cooper pairs \cite{Frigeri2004, Fak2014}. 

A direct way to gain insight in non-phonon driven superconductivity is to study the coupling of magnetic order and superconductivity. Magnetic superconductors feature a variety of different behavior when tuned via external parameters, such as pressure, chemical substitution or magnetic fields \cite{Norman2011, Pfleiderer2009, White2015}. In most materials a competition between both phenomena is observed \cite{Norman2011, Pfleiderer2009, White2015}, but there also exist cases in which magnetic order and superconductivity cooperate \cite{Huxley2015, Kenzelmann2008, Kenzelmann2017}. Examples include the heavy-fermion compound UGe$_2$, where superconductivity is only stable in the presence of ferromagnetic order \cite{Huxley2015}, or CeCoIn$_5$ where magnetism  only appears inside the superconducting phase \cite{Kenzelmann2008, Kenzelmann2017}. The latter phenomena has been discussed theoretically already for some time \cite{Fulde1964, Larkin1965, Psaltakis1983, Shimahara2000, Lebed2006, Aperis2008, Yanase2009, Kato2011, Suzuki2011, Agterberg2009, Michal2011, Mineev2016, Kim20172, Hatakeyama2015, Hosoya2017}, but it remains an open question how magnetic order can emerge from superconductivity. 

The series Nd$_{1-x}$Ce$_x$CoIn$_5$ reveals a competition between static magnetic order and superconductivity for $x$ $>$ 0.78 at zero field \cite{Hu2008}. 5\% Nd doped CeCoIn$_5$, however, features an auxiliary field-induced magnetic phase (Q-phase) that is only stable within the superconducting condensate and that collapses in a first-order transition at the upper critical field \cite{Mazzone2017}. This behavior provides evidence for a cooperative magneto-superconducting ground state in the Q-phase with a coupling that is similar to the one of undoped CeCoIn$_5$.  The high-field Q-phase of 5\% Nd doped CeCoIn$_5$ is separated from its low-field SDW-phase via a quantum phase transition at $\mu_0H^*\approx$ 8 T \cite{Mazzone2017}. Both magnetic phases feature the same amplitude modulated spin-density wave (SDW) order with an ordered magnetic moment, $\mu\approx$ 0.15$\mu_B$, that is oriented along the $c$-axis. The magnetic propagation vector, \textbf{Q}$_{1,2}$ = ($q$, $\pm q$, 0.5) with $q\approx$  0.445, is similar to the one of the Q-phase in CeCoIn$_5$ and directed along the nodal direction of the $d_{x^2-y^2}$-superconducting order parameter \cite{Mazzone2017}. However, it is currently an open question how the symmetry of superconductivity is affected in the Q-phase.
\begin{figure*}[tbh]
\includegraphics[width=\linewidth]{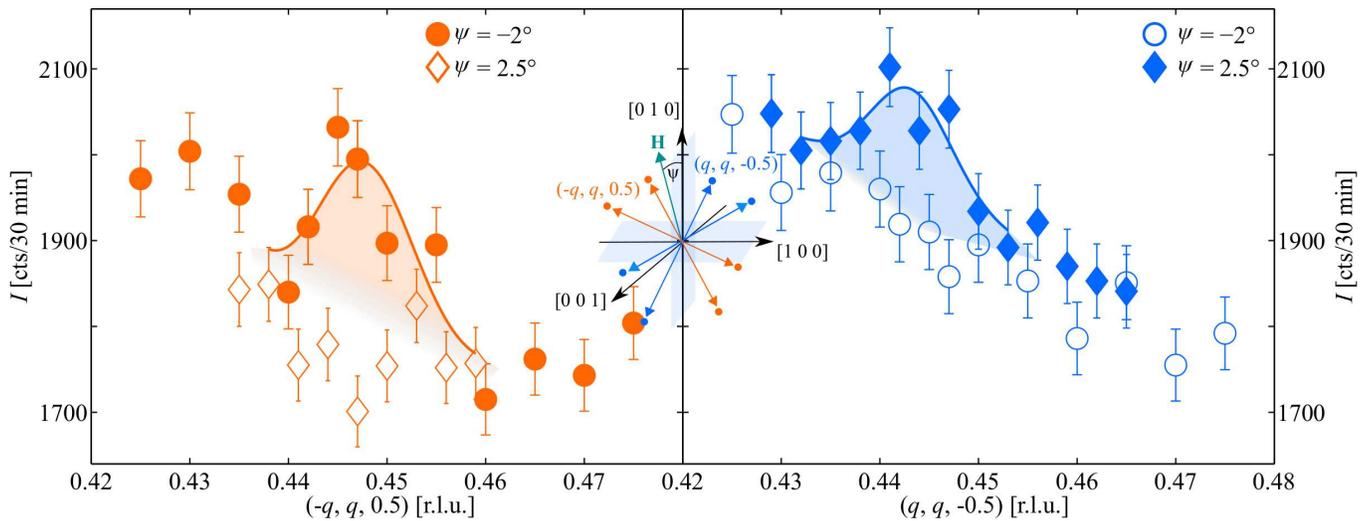}
\caption{\textbf{Switching in the Q-phase.} Diffracted neutron intensity along the tetragonal plane of representatives of the two magnetic domains, \textbf{Q$_1$} = ($q$, $q$, 0.5) and \textbf{Q$_2$} = ($q$, -$q$, 0.5), measured at $\mu_0H$ = 10.4 T and $T$ = 40 mK for $\psi$ = 2.5 and -2$^\circ$. Solid lines represent Gaussian fits on a sloping background.}
\label{fig1}
\end{figure*}

Here, we study the population of the two magnetic domains, \textbf{Q}$_{1,2}$, in the SDW- and Q-phase of Nd$_{0.05}$Ce$_{0.95}$CoIn$_5$ by means of neutron diffraction for magnetic fields oriented close to \textbf{H}$||$[0 1 0] and along the [1 $\bar{1}$ 0]-direction in reciprocal lattice units (r.l.u.). We find a field-induced redistribution of the domain-population for \textbf{H}$||$[1 $\bar{1}$ 0], where the intensity in \textbf{Q}$_{2}$ is suppressed at $\mu_0H_d$ =  3.6(6) T. The Q-phase features a single spin-density modulation direction except for a small field range of $\pm$2.5$^\circ$ around \textbf{H}$||$[0 1 0], where a continuous crossover of populated domains is observed. The behavior is different in the SDW-phase, where the magnetic domains remain equally populated for magnetic fields close to the $a$-axis. 

\section*{Results}

Fig. \ref{fig1} displays two magnetic Bragg peaks that belong to the magnetic domains \textbf{Q$_1$} and \textbf{Q$_2$}. The diffracted neutron intensity was measured for wave-vector transfers, ($\pm q$, $q$, $\mp$0.5), along the tetragonal plane in the Q-phase of Nd$_{0.05}$Ce$_{0.95}$CoIn$_5$. The magnetic field equals $\mu_0H$ = 10.4 T and is oriented along $\psi$ = 2.5 and -2$^\circ$ relative to the $a$-axis. The neutron diffraction results show a populated \textbf{Q$_1$}-domain for $\psi$ = 2.5$^\circ$, while \textbf{Q$_2$} is suppressed. When rotating the field to $\psi$ = -2$^\circ$ the domain population is switched. A tunable mono-domain state is also found in the Q-phase of CeCoIn$_5$ \cite{Gerber2014}. 

\begin{figure}[tbh]
	\begin{center}
\includegraphics[width=\linewidth]{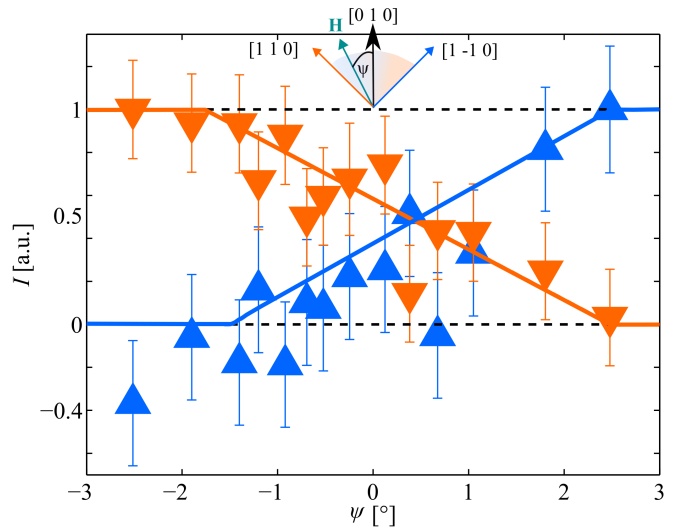}
\caption{\textbf{Domain population.} Background subtracted peak intensity as a function of $\psi$ for representatives of \textbf{Q$_2$} in orange and \textbf{Q$_1$} in blue measured at $\mu_0H$ = 10.4 T. Solid lines are guide lines to the eyes. Dashed line represent the fully suppressed and fully populated domains.}
\label{fig2}
	\end{center}
\end{figure}

Fig. \ref{fig2} depicts the angular dependence of the background subtracted peak intensity of both domains inside the Q-phase at $\mu_0H$ = 10.4 T. When rotating the magnetic field through [0 1 0] one domain continuously depopulates while the other one is populated. The neutron diffraction results reveal a crossover region of $\Delta\psi\approx$ 5$^\circ$ where both domains are at least partly populated. Although this is much broader than in undoped CeCoIn$_5$, this is relatively sharp considering that the Q-domain must be pinned to the Nd dopants \cite{Gerber2014}. 

\begin{figure}[tbh]
\includegraphics[width=\linewidth]{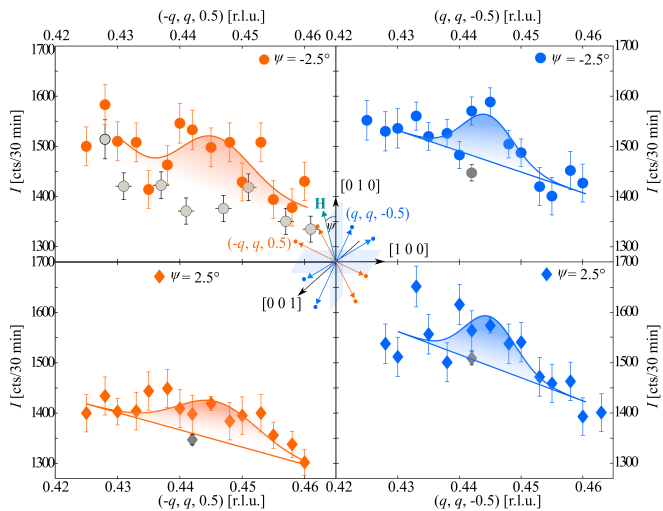}
\caption{\textbf{Switching in the SDW-phase.} Diffracted neutron intensity along the tetragonal plane of \textbf{Q$_2$} in orange and \textbf{Q$_1$} in blue, measured at $\mu_0H$ = 2 T and $T$ = 40 mK for $\psi$ = $\pm$2.5$^\circ$. Background in gray was measured at $\mu_0H$ = 11.9 T. Solid lines represent Gaussian fits on a sloping background.} 
\label{fig3}
\end{figure}

We further studied the magnetic domain population in the SDW-phase. Diffracted neutron intensity along the tetragonal plane of \textbf{Q$_1$} and \textbf{Q$_2$} at $\mu_0H$ = 2 T is shown in Fig. \ref{fig3}. The gray dots denote the background that was measured at $\mu_0H$ = 11.9 T $>$ $\mu_0H_{c_2}$.  Within the SDW-phase no change in the domain population is observed in the vicinity of \textbf{H}$||$[0 1 0]. This is in strong contrast to the Q-phase, where only one of both domains is populated for $\psi$ = $\pm$2.5$^\circ$. 

\begin{figure}[tbh]
	\begin{center}
\includegraphics[width=\linewidth ]{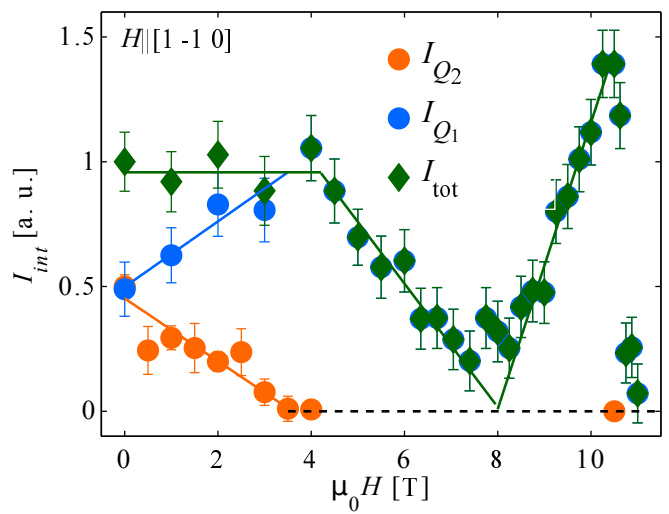}
\caption{\textbf{Field-induced Domain imbalance.} Field dependence of the two magnetic domains, \textbf{Q}$_1$ and \textbf{Q}$_2$, in Nd$_{0.05}$Ce$_{0.95}$CoIn$_5$ for \textbf{H}$||$[1 $\bar{1}$ 0] and the total intensity $I_{tot}$ = $I_{Q_1}$ + $I_{Q_2}$. The curve is reconstructed from the data presented here and the ones reported in Ref. \cite{Mazzone2017}. The Solid lines represent weighted linear fits. }
\label{fig4}
	\end{center}
\end{figure}

The field dependence of the integrated Bragg peak intensity at (0.56, 0.44, 0.5) is represented by orange circles in Fig. \ref{fig4} for \textbf{H}$||$[1 $\bar{1}$ 0]. It demonstrates that the magnetic \textbf{Q}$_2$-domain gradually weakens with increasing magnetic field and is suppressed at $\mu_0H_d$ = 3.6(6) T. This is in strong contrast to the field dependence of the \textbf{Q}$_1$-domain, whose intensity increases at small fields and reveals a broad maximum around $\mu_0H\approx$ 4 T  \cite{Mazzone2017}.

In addition, Fig. \ref{fig4} compares the scaled, integrated intensity of  the two magnetic domains, $I_{Q_1}$ and $I_{Q_2}$, with the total integrated intensity $I_{tot}$ = $I_{Q_1}$ + $I_{Q_2}$ for \textbf{H}$||$[1 $\bar{1}$ 0]. This plot combines new measurements with those published earlier \cite{Mazzone2017}. The normalization of the integrated intensity was chosen to have equal population at zero field, respecting the tetragonal symmetry.

The field dependence of $I_{Q_1}$ and $I_{Q_2}$ show that fields smaller than $H_d$ trigger a redistribution of the domain population. The intensity in \textbf{Q}$_1$ is enhanced, while the magnetic domain in the plane along the field \textbf{Q}$_2$ is reduced, such that the total integrated intensity remains constant. Increasing the field further yields a mono-domain state, where only the \textbf{Q}$_1$-domain is populated.

\section*{Discussion}

5\% Nd substituted CeCoIn$_5$ features magnetic order with a moment orientation along the tetragonal $c$-axis \cite{Mazzone2017}. The Zeeman coupling (\textbf{M}\textbf{H}) vanishes for fields applied in the tetragonal basal plane and cannot drive the field and angular dependent population of \textbf{Q}$_1$ and \textbf{Q}$_2$. 

An anisotropic spin susceptibility can originate from spin-orbit interactions as observed, for instance, in the non-centrosymmetric superconductor CePt$_3$Si \cite{Frigeri2004, Fak2014}. It has been suggested that a similar phenomena can appear in a multiband metal with tetragonal symmetry, such as CeCoIn$_5$, when the spin-orbit coupling reduces the space group symmetry by a basal in-plane field \cite{Mineev2016}. Based on this microscopic theory , Kim $et.$ $al.$ developed a phenomenological Landau model, which includes a weak spin-orbit coupling term in its free energy density \cite{Kim20172}. It considers a magnetic moment arrangement perpendicular to the basal plane, as found in the two antiferromagnetic phases of Nd$_{0.05}$Ce$_{0.95}$CoIn$_5$. Depending on the Landau parameters, the model predicts either (A) the coexistence of the Q-domains whose population is tuned with the field direction, or (B) the presence of only one Q-domain with a sharp switching \cite{Kim20172}. 

Scenario A yields an equal domain population for $\psi$ = 0$^\circ$ and a maximal difference for $\psi$ = 45$^\circ$. This is consistent with the neutron diffraction results at $\mu_0H$ = 2 T, where two populated domains are found that feature a comparable intensity for \textbf{H}$||$[0 1 0] and a 80\% suppression of \textbf{Q}$_2$ for \textbf{H}$||$[1 $\bar{1}$ 0]. Our data in the SDW phase can thus be explained with spin-orbit couplings mediating a field-induced repopulation of the Q-domains. The much sharper switching at high fields, however, is not consistent with scenario A and suggest qualitatively different behavior more consistent with scenario B. This means that the Landau theory by Kim $et.$ $al.$ \cite{Kim20172} can describe the SDW and Q-phase separately, but not both phases simultaneously, and different phenomenological parameters would be needed for the low- and high-field phases. This points towards a qualitative change of the electronic structure at $H^*$. 

Alternatively it has been suggested that $d$-wave superconductivity and magnetic order is coupled via a spatially modulated spin-triplet superconducting order parameter \cite{Agterberg2009, Hosoya2017}. The formation of triplet superconductivity also relies on spin-orbit interactions and the additional order parameter supports a sharp switch of the modulation direction for a field around the $a$-axis \cite{Agterberg2009, Gerber2014}. A supplementary superconducting gap is consistent with a non-magnetic primary order parameter in the Q-phase that is postulated by the identical magnetic symmetry in the SDW- and the Q-phase \cite{Mazzone2017}. Such a scenario can also account for the thermal conductivity results of the Q-phase in CeCoIn$_5$ that reveal a reduced quasiparticle excitation spectrum perpendicular to the populated \textbf{Q}-domain \cite{Kim2016}. A symmetry analysis suggests that the two modulation directions belong to different irreducible representations and that a $p$-wave order parameter is aligned along the suppressed domain \cite{Gerber2014, Kim2016}. As a result, the switching has to be sharp, albeit it may be broadened by pinning to Nd-ions.

It has been suggested that the Q-phase arises from the condensation of a superconducting exciton that creates a novel superconducting state \cite{Michal2011}. The superconducting spin resonance that is observed in CeCoIn$_5$ at zero field appears at the same wave-vector as static magnetic order in the Q-phase \cite{Raymond2015}. Under magnetic field the resonant splits and the lowest mode may condense into the ground state at the Q-phase boundary \cite{Michal2011, Stock2012, Raymond2012, Akbari2012}. In 5\% Nd doped CeCoIn$_5$ the resonance is not affected by the onset of static magnetic order at zero field, which provides evidence for a decoupling of these fluctuations from magnetic order in the SDW-phase \cite{Mazzone20172}.

In summary, we demonstrate that an in-plane rotation of the magnetic field in Nd$_{0.05}$Ce$_{0.95}$CoIn$_5$ triggers a magnetic domain imbalance that is distinct in the SDW- and the Q-phase. We find two domains with no preferable spin-density modulation direction in the low-field SDW-phase, for fields close to the $a$-axis. At low fields, a field-induced change of the relative domain population is observed when the field is applied along the diagonal direction of the tetragonal plane.  The selection of a mono-domain state becomes relatively sharp in the high-field Q-phase, where the spin-density modulation direction can be switched by a few-degree rotation of the field around the $a$-axis. The low- and high-field behavior cannot be simultaneously explained by the available phenomenological theories, and requires a modification in the coupling between superconductivity and magnetic order at $H^*$. We suggest that an additional superconducting order parameter of $p$-wave symmetry emerges at high fields and intertwines magnetic order with $d$-wave superconductivity.

\section*{Methods}

The neutron diffraction experiments at $T$ = 40 mK and up to $\mu_0H$ = 11.9 T were carried out on the thermal neutron lifting-counter two-axis spectrometer D23 at the Institut Laue Langevin, Grenoble France. The single crystal ($m$ = 64 mg) was placed in a vertical-field magnet with dilution insert and exposed to an incident neutron wavelength of $\lambda$ = 1.27 \AA. The crystal was oriented either with the vertical axis parallel to [1 $\bar{1}$ 0] or along [0 1 0]. In the latter setup the tetragonal $a$-axis was tilted into the basal plane using a non-magnetic piezoelectric sample rotator (type ANGt50 from attocube system AG) inside the dilution refrigerator (see Supplementary Materials of Ref. \cite{Gerber2014}).  The relative field direction was directly measured via the vertical tilt of the structural (2, 0, 0) Bragg peak. The relative angle between the magnetic field direction \textbf{H} and the tetragonal $a$-axis, [0 1 0], is denoted as $\psi$

The solid lines in Fig. \ref{fig1} and \ref{fig3} represent Gaussian fits to the neutron diffraction data. Magnetic Bragg peaks at ($q$, $q$, -0.5) were fitted using a Gaussian line shape on a linear background. The background at  (-$q$, $q$, 0.5) and $\psi$ $\leq$ -2$^\circ$ could not be described satisfactorily by a linear behavior solely. We used an additional Gaussian component centered at $q$ = 0.427. The width of the magnetic Bragg peaks in each domain was fixed to the mean value of all corresponding fits.

The integrated intensity at (0.56, 0.44, 0.5) for $\mu_0H$ = 0, 1,  2, 3 and  4 T and $T$ = 40 mK along  \textbf{H}$||$[1 $\bar{1}$ 0] was obtained from scans along ($q$ + $h$, $q$, 0.5), where $h$ was chosen such that the scan was centered at (1, 0, 1) -  \textbf{Q$_2$}. The integrated intensity at $\mu_0H$ = 0.5, 1.5, 2.5 and 3.5 T was determined from the background subtracted Bragg peak intensity multiplied with the averaged peak width found in the $q$-scans. A weighted linear fit to these results yields $\mu_0H_d$ = 3.6(6) T.

All data needed to evaluate the conclusions in the paper are present in the paper. Additional data available from authors upon request. Correspondence and requests for materials should be addressed to D. G. M. (email: danimazzone@gmail.com)
\def\bibsection{\section*{\refname}}

\section*{Acknowledgements}

We acknowledge the Institut Laue-Langevin for the allocated beam time on D23. We thank P. Fouilloux for technical assistance. Discussions with Duk Young Kim and Roman Movshovich are acknowledged. In addition, we thank the Swiss National Foundation (grant No. 200021\_147071, 206021\_139082 and 200021\_138018). This work was also supported by the Swiss State Secretariat for Education, Research and Innovation (SERI) through a CRG-grant.

\section*{Author contributions statement}
D. G. M., S. R., J. L. G and M. K. planed and led the project. The experiments were carried out by D. G. M., R. Y., M. B., J. L. G., S. R., E. R., and M. K. The sample was grown by G. L and the data analyzed by D. G. M. The manuscript was written by D. G. M., S. R., J. L. G. and M. K. with the input of all co-authors.

\section*{Additional information}

\textbf{Competing financial interests}: The authors declare that they have no competing interests. 

\end{document}